# Quantum theory is classical mechanics with non-local existence

John Hegseth
Department of Physics
University of New Orleans
New Orleans, LA 70148[*]

**Abstract**

I propose a new and direct connection between classical mechanics and quantum mechanics where I derive the quantum mechanical propagator from a variational principle. This variational principle is Hamilton's modified principle generalized to allow many paths due to the non-local existence of particles in phase space. This principle allows a physical system to evolve non-locally in phase space while still allowing a representation that uses many classical paths. Whereas a point in phase space represents a classical system's state, I represent the state of a non-local system by a mixed trajectory. This formulation naturally leads to the transactional interpretation for resolving the paradoxes of the measurement problem. This principle also suggests a more flexible framework for formulating theories based on invariant actions and provides a single conceptual framework for discussing many areas of science.

**Introduction**

Physical laws expressed as variational principles are appealing for their simplicity and generality [1], [2], [3]. In addition, they often provide information about a system's stability based on the type of extrema (e.g., minima, maxima, etc.) [4]. In classical physics the type of extrema depends on the details of a specific system's Hamiltonian [2], [5]. In classical Hamiltonian dynamics the action $S[p(t),q(t),t] = \int (p\dot{q} - H(p,q))dt$ is extremized using the modified Hamilton's principle. Another action $R[p,q,t] = \int (-q\dot{p} - H(p,q))dt$ may also be used to derive Hamilton's equations or a Hamilton-Jacobi equation [5]. The classical dynamics are time reversal invariant and both $R$ and $S$ require perfectly known beginning ($t_i$) and ending ($t_f$) conditions for extremization: $\delta p(t_i)=0$, $\delta p(t_f)=0$ and $\delta q(t_i)=0$, $\delta q(t_f)=0$ respectively. The extrema completely determines a trajectory as a path in phase space (e.g., with *N* particles a path is in *6N* dimensional phase space ($p_1, p_2,..., p_N$ ; $q_1, q_2, ..., q_N$) ). This path is completely determined in the past and the future so that any given point on the trajectory is necessary and sufficient to completely specify the system's state. I also note that these modified Hamilton's Principles are easily expressed in a manifestly covariant form, i.e., with an invariant parameter $\lambda$ (e.g., the proper time) $S[\vec{p}(\lambda),\vec{q}(\lambda),\lambda] = \int (\vec{p} \bullet \vec{q}')d\lambda$ and $R[\vec{p}(\lambda),\vec{q}(\lambda),\lambda] = -\int (\vec{p}' \bullet \vec{q})d\lambda$ where $\vec{p}(\lambda) = (H/c, p_1, p_2, p_3)$, $\vec{q}(\lambda) = (ct, q_1, q_2, q_3)$, $\vec{p}' = \frac{d\vec{p}}{d\lambda}$, $\vec{q}' = \frac{d\vec{q}}{d\lambda}$, and "$\bullet$" includes the metric.

In this paper I propose another variational principle that describes the motion when perfect information about *p* and *q* does not exist. This is done by taking $S = S(q,\dot{q},t)$, $R = R(p,\dot{p},t)$, and generalizing them to include all possible paths. I then introduce two mixed path distribution functionals over the possible *p(t)* paths *α[p(t)]* and the possible *q(t)* paths *β[q(t)]*. I will first discuss how the mixed path distributions accommodate non-local phase space existence. I next discuss how they are normalized for particle translation in space-time and for internal variables (fermions and bosons). I finally show how Hamilton's principle is generalized with mixed path distributions and how the probability amplitude naturally follows.

**The non-locality of reality**

Well-known experimental evidence and theoretical arguments suggest that quantum mechanical objects are non-local in space and time [6], [7], [8]. In this approach, I take the minimum of the Heisenberg Uncertainly Principle (HUP, $\delta p \delta q \geq h$ and $\delta E \delta t \geq h$) as the essential statement of this non-local existence. i.e., a particle's empirical existence in phase space is an incompressible area given by $\delta p \delta q = h$ ($\delta E \delta t = h$) for each conjugate pair of variables. As I will later show, for compact internal variables this area may be infinitely deformable for bosons or rigid for fermions. Just as a classical phase space point evolves on a trajectory that extremizes the modified Hamilton's Principle, a particle's possible phase space trajectories evolve such as to extremize a generalized version of the modified Hamilton's Principle. Whereas a point in phase space represents a classical system's state, the state of a system with non-local phase space existence is represented by a mixed trajectory. Because the modified Hamilton's Principle is easily generalized to be manifestly covariant, this discussion may be directly generalized for relativistic invariance. This discussion may also be directly generalized to many particles or fields (see, e.g., Reference [9]) but I will only discuss non-relativistic particles.

Let experimenters *A* and *B* observe a system of particles. As discussed above, it is not possible to determine a phase space trajectory because specific phase space points do not empirically exist. We may still construct a continuous position space or momentum space, because this is not excluded by HUP. This implies that there are many possible paths in either position space or momentum space. Let *A* empirically sample the momentum path *p(t)* (i.e., $\delta q \to \infty$ for *A*) and let *B* only measure points on the position path *q(t)* (i.e., $\delta p \to \infty$ for *B*). Between sampling time interval $\Delta t$, *A* may observe an initial and final *p* and extremize *R* where *R* must be expressed in terms of *p*, $\dot{p}$, and *t*, $R = R(p, \dot{p}, t)$. During $\Delta t$, *B* may observe an initial and final *q* and extremize *S* ($S = S(q, \dot{q}, t)$). Figure 1 shows a simplified example with a simplified uncertainty principle during two time intervals. In this example $q_i$ can be localized uniquely at any time but there are three possible *p* values for each *q* that *B* could observe, i.e., there is an uncertainty principle of three possible *p*'s per definite *q* at each time.

If a particle's initial *q* is perfectly localized, HUP implies that all *p*'s are possible at $t_i$ (the initial phase space area of the particle is deformed into a line, c.f., Figure 1). In addition, at the next time interval all *q*'s are possible, i.e., the phase space area has expanded. The position is non-local an infinitesimal time later consistent with known experiments. The HUP has two parts to the "reality" of particle non-locality, an ontological part as discussed above $\delta p \delta q = h$, and an epistemological part $\delta p \delta q > h$ that could be compressed through a localizing event. The EPR view of "reality" corresponds to an ideal phase space point as a particle descriptor, whereas I use a broader view of the reality of a particle or system as a phase space region [8]. A predictable linear macroscopic system has only insignificant ontological uncertainty - effectively none. An unpredictable macroscopic system, however, such as a chaotic oscillator or a turbulent fluid, that is sensitive to initial conditions and amplifies initial uncertainties will initially exhibit mostly epistemological uncertainty, from e.g., thermal fluctuations. If given sufficient time, however, such a system will also exhibit ontological uncertainty. This implies that the slightest particle interaction in a multi-particle closed system will ultimately result in ergodicity (note that the perfectly reflecting walls of an ideal gas container is a significant interaction). The "ergodic hypothesis" naturally follows from ontological uncertainty, i.e., when an initial uncertainty of $\delta p \delta q = h$ is amplified to cover the system's possible phase space, then ergodicity is assured.

**Extreme non-locality and time reversal**

In principle, there is an upper limit to the possible energy and momentum ($E/c, p$) in a finite universe. A finite universe implies a maximum ($E_{max}, p_{max}$). This maximum is also the maximum value of the particle's ($|\delta E|, |\delta p|$) where it is possible (however unlikely) that the particle has all of the energy and momentum in the universe. This corresponds to a natural cutoff for the regularization of quantum field theory and implies a lower limit of localization in ($t, q$) as is often assumed in quantum gravity [10]. A similar upper limit also applies to ($t, q$), where an upper limit ($t_{max}, q_{max}$) implies a lower limit in localization of ($|\delta E|, |\delta p|$). Although this implies that a particle cannot empirically exist at a point in space-time (or momentum-energy), I will sidestep this cosmological issue for now, and take this to be an effective point compared to the other scales in the problem, i.e., I take $\delta q \to 0, \delta p \to \infty$; $\delta t \to 0$, $\delta E \to \infty$ (or vice versa) as relative or effective limits. I note that these limits justify continuous space-time (or momentum-energy) to be used in this discussion for $A$ or $B$'s particle observation. This also shows the validity of an effective field theory using a space-time domain (or a momentum-energy domain) in a covariant formulation.

An uncertainly principle clearly allows a very significant increase in the number of possible paths in time as illustrated in Figure 1. If $B$ fixed an endpoint at a later time $t_f$, i.e., the fixed endpoint $q(t_f)$, the number of possibilities would be reduced. Although it is clear how the HUP can generate more possible paths, this sudden localization of $q(t_f)$ seems quite counter-intuitive, much like violating the second law or the "collapse of the wavefunction". In fact, $q(t_f)$ is a boundary condition enforced by the experiment (a localizing event in $q$ leaving $p$ de-localized) at a known later time $t_f > t_i$. I note that the HUP $\delta E \delta t \geq h$ also suggests an indefinite existence in time, i.e., for relatively definite energy, many possible sequences of events exist. In fact, in an infinitesimal time interval between events, the final and initial times could be reversed with a corresponding negative $\delta E$. It follows that paths may evolve forward or backward in time. We may then preserve time reversal symmetry by evolving $A$'s and $B$'s paths forward and backward in time. The evolution is non-local and a-temporal, consistent with Bell's theorem and delayed choice experiments [7]. I note that evolving a given path both forward and backward in time is redundant and cannot be empirically distinguished from a consistently evolving path between the boundary points, i.e., distinguishable paths either travel forward or backward in time as shown in Figure 1. Paths are separated into two classes; forward evolving and backward evolving. Causality is preserved through the boundary conditions *δq(t_i)=0, δq(t_f)=0* that break time reversal symmetry. If, however, $t_f - t_i < h/E_{max}$, then time reversal symmetry cannot be broken because time ordering does not empirically exist. These lower limits of localization suggest that empirical observation is limited and therefore space-time – in the empirical sense – does not exist at smaller scales. As suggested above, I may still discuss these limits or "uncertainties" in terms of an ideal space where points and a continuum etc. may be hypothesized to exist. This is similar to discussing a finite empirical universe in terms of an ideal space where coordinates extend to infinity.

**Abnormal distributions**

I next define the distributions for these possible ideal paths: *α[p(t)]* for *A* and *β[q(t)]* for *B* between fixed beginning and ending points. Below I will discuss how these distributions are selected to extremize a generalized form of Hamilton's principle. It is important to note that for each possible complete ideal phase space path (*p(t), q(t)*) there is a one-to-one relation between *q(t)* and *p(t)* (below I will not refer to *p(t)* or *q(t)* as ideal unless the meaning is possibly ambiguous). For each possible *q(t)* and $\dot{q}(t)$ in *B*'s set of possible paths, he can infer a possible *p(t)* using Hamilton's equation $\dot{q} = H_p$. Similarly, for each possible *p(t)* and $\dot{p}(t)$ in *A*'s set, he can infer a *q(t)* path using $\dot{p} = -H_q$. I note that this is a one-to-one relation between *possible* paths in forward time and backward time. *α[p(t)]* and *β[q(t)]* are independent distributions, however, because (*p(t)*, $\dot{p}(t)$) and (*q*, $\dot{q}(t)$) are independent sets of ideal variables for the description of a particle's behavior derived from independent variational principles [5]. Because both *A* and *B* could measure or infer *q(t)*, the probability for a given *q(t)* is *β[q]β[q]= β²[q]*. Similarly, the probability for a given *p(t)*, that could be measured by *A* and *B*, is *α[p]α[p]= α²[p]*. The probability $P_p$ for a particular phase space path between specified points at $t_i$ and $t_f$, e.g., ($q_i$, $t_i$) and ($q_f$, $t_f$), is the probability that *A* and *B* could observe the path in terms of *q(t)* or that *A* and *B* could observe the path in terms of *p(t)*, i.e., $P_p$ = *β[q]β[q]* + *α[p]α[p]*. I note that *α[p]* or *β[q]* may be negative and are not probabilities. I call the distributions *–1≤β≤1* and *-1≤α≤1* mixed paths because, as I show later, they are analogous to mixed strategies in the Game Theory of competitive behavior [12]. *α²* (or *β²*) is the probability that *A* and *B* could observe a given *p(t)* (or *q(t)*).

I now consider many possible ideal phase space paths for particles. A similar argument can also be made for the internal variables of bosons and fermions. Let $B$ hypothesize a given path, say $q_1$. $A$ could measure any of the possible paths $q_1, q_2, ..., q_n$ where this finite set is defined by discrete points in space-time as given in reference [13]. The probability that $B$ could observe $q_1$ while $A$ observes any of the paths in terms of $q$ is $\beta[q_1]\Sigma_j\beta[q_j]$ for $j=1,2 ...,n$. The probability for $A$ and $B$ to observe any of the paths in terms of $q$ is $\Sigma_k\beta[q_k]\Sigma_j\beta[q_j] = \{\Sigma_j\beta[q_j]\}^2$. Similarly, the probability for $A$ and $B$ to observe any of the paths in terms of $p$ is $\{\Sigma_j\alpha[p_j]\}^2$. The total probability $P_T$ to observe any of the possible phase space paths between given beginning and ending spatial points is $P_T(q_i,q_f)=\{\Sigma_j\alpha[p_j]\}^2+\{\Sigma_j\beta[q_j]\}^2$. In terms of the relative probability with respect to $P_T$, I get the normalization $\{\Sigma_j\alpha[p_j]\}^2+\{\Sigma_j\beta[q_j]\}^2=1$. I can now allow a fan out between paths at any instant and any position or momentum in an interval $T=n\Delta t$ by letting $\Delta t\to 0$ and $n\to\infty$ to define a path integral $\Sigma_j\alpha[p_j]\to \int Dp\alpha[p(t)]$ and $\Sigma_j\beta[q_j]\to \int Dq\beta[q(t)]$, where $Dq$ and $Dp$ are the effective measures of the integral [13]. I note that these "measures" may be thought of as a shorthand notation for these large but finite sums as these are effective limits as discussed above. I can find the probability $P_T$ for a particle to start at $q_i$ and end at $q_f$ by summing over all possible paths or performing the path integrals $\int Dp\alpha[p(t)]$ and $\int Dq\beta[q(t)])$ to get $P_T(q_i,q_f) = \{\int Dp\alpha[p(t)]\}^2 +\{\int Dq\beta[q(t)]\}^2$.

$A$ and $B$ may measure their respective *internal* variables if the particle has integer spin (e.g., $B$ is capable of measuring the angle variable using polarizing filters). The same normalization as above follows. If a particle has an internal ½ integer spin degree of freedom, however, as discussed in reference [5], then only $A$ is able to actually measure $p(t)$ (e.g., measure $p=\pm constant$ through an electron's magnetic moment). The phase space (empirical) existence is a (ideal) "line" $\delta q\to\infty$, $\delta p\to 0$; $\delta t\to\infty$, $\delta E\to 0$ of area $h$ that cannot deform. Note that in this case $q$ is a compact angle variable and $\delta q$ is not constrained by an upper limit whereas $\delta t$ and $\delta E$ are constrained by effective limits. Because $p=\pm const.$, any given $p(t)$ path has discontinuities and $A$ cannot invert $\dot p = - H_q$ to find a corresponding $q(t)$. A specific $q(t)$ cannot be inferred from a specific $p(t)$. The distributions $\alpha[p(t)]$ and $\beta[q(t)]$ may be defined between $(p_i, p_f)$ but not between the angle variables $(q_i, q_f)$ that cannot be fixed in advance as they are impossible to observe because of a complete ontological uncertainty. A given $q(t)$ can still be hypothesized and $\beta[q(t)]$ still exists as it is defined over all possible $q(t)$. Because $A$ could not measure or infer $q(t)$, if he could measure $p(t)$, the probability for a given $q(t)$ that could be measured by $A$ and $B$ is $\beta^2[q] = 0$. In other words, it is not possible for $A$ and $B$ to measure a given $q(t)$. In addition, because a possible $q(t)$ would have cusps corresponding to $p(t)$ discontinuities, $B$ can't invert $\dot{q} = H_p$ to get a one-to-one relations between $q(t)$ and $p(t)$. Because $B$ could not infer a unique $p(t)$ from a given $q(t)$, the probability for a given $p(t)$, measured by $A$ and $B$, is $\alpha[p]\alpha[p] = \alpha^2[p]=0$. Although an ideal $(p, q)$ space can still be defined, I cannot associate a specific empirical $q(t)$ to a $p(t)$ or vice versa. Separated $p(t)$ and $q(t)$ paths may be constructed and the corresponding distributions $\alpha[p]$ and $\beta[q]$ must not be zero. They must be anti-commuting Grassman numbers: $\alpha[p]\beta[q]=-\beta[q]\alpha[p]$. The probability $P_p$ for a particular path between $(p_i, t_i)$ and $(p_f, t_f)$, is no longer $\beta^2 + \alpha^2$, i.e., $\beta[q]\beta[q] + \alpha[p]\alpha[p]=0$. In fact, this is a normalization statement that such an observation process is impossible. It is possible, however, for $A$ to represent a path in terms of $p(t)$ and $B$ to represent a path in terms of $q(t)$, i.e., $\alpha[p]\beta[q]\neq 0$. It is impossible for $A$ and $B$ to represent the path in terms of either $p(t)$ and $q(t)$ or $q(t)$ and $p(t)$: $\alpha[p]\beta[q] + \beta[q]\alpha[p] = 0$, i.e., $A$ and $B$ can't know both $p(t)$ and $q(t)$ of a given ideal path. We may then define a positive probability, such as $\alpha[p]\beta[q]$, as the probability of a possible outcome as usual. The negative quantity $\beta[q]\alpha[p]$, is not just impossible but anti-possible, i.e., it is an independent alternative that annihilates a possible outcome to produce a normalization statement of absolute certainty of impossibility. These events together are impossible because of the broken classical connection between $p$ and $q$ that results from the absence of possible information about $q$ and the allowed discontinuous evolution of $p(t)$ and $\dot{q}(t)$. I note that the previous case could also have discontinuous evolution. This case, however, does not have any scale separation and the discontinuities cannot be ignored through a scaling or renormalization argument. By generalizing the mixed path to include Grassman numbers, the anti-commuting behavior of fermions is described. $\alpha\beta$ is the probability for a given path and $P_T(p_i,p_f) = \int Dp \int Dq \alpha[p(t)]\beta[q(t)]$. The two possible values of $p=\pm const.$ at any given time implies that $\alpha[p]$ and $\beta[q]$ can be represented by spinors [5] or with relativistic invariance as Dirac spinors.

**A piece of the action**

I now show the generalization of Hamilton's principle that governs the form of the actual or "best" pair of mixed paths $\alpha_0[p]$ and $\beta_0[q]$. Because any ideal path, indexed by $j$ for $A$ and $k$ for $B$, can be represented in terms of $p(t)$ or $q(t)$, there is the same number $n$ of possible $p$ paths and $q$ paths in a set of $n$ ideal phase space paths, as shown in Figure 2 for three paths. Corresponding to each $p_j(t)$ or $q_k(t)$ is $\alpha[p_j]=\alpha_j$ and $\beta[q_k]=\beta_k$. I represent these mixed paths with the vectors $\boldsymbol{\alpha}=(\alpha_1, \alpha_2, ..., \alpha_n)$ and $\boldsymbol{\beta}=(\beta_1, \beta_2, ..., \beta_n)$. I can also define a square matrix for the action $S_{jk}$ corresponding to all possible actions for the $n$ ideal paths with given $q_i$ and $q_f$. The elements of this action matrix $S_{jk}$ are real values of the action evaluated at paths $p_j$ and $q_k$, i.e., $S[p_j, q_k]=S_{jk}$. In the three-path case, shown in Figure 2, the action matrix $\underline{S}$ is written

$$\underline{S} = \begin{Vmatrix} S_{11} & S_{12} & S_{13} \\ S_{21} & S_{22} & S_{23} \\ S_{31} & S_{32} & S_{33} \end{Vmatrix}.$$

I combine $\boldsymbol{\alpha}$, $\boldsymbol{\beta}$, and $S_{jk}$ to form the generalized action $\boldsymbol{\alpha}^T\underline{S}\boldsymbol{\beta}$ where $\boldsymbol{\alpha}^T$ is the transpose of $\boldsymbol{\alpha}$. In analogy to the case with perfect information, I extremize this generalized action by finding optimal mixed paths $\boldsymbol{\alpha}_0$ and $\boldsymbol{\beta}_0$. $A$ selects $\boldsymbol{\alpha}_0$ so as to extremize $\boldsymbol{\alpha}_0^T\underline{S}\boldsymbol{\beta}$ for any given $\boldsymbol{\beta}$, and $B$ selects $\boldsymbol{\beta}_0$ to extremize $\boldsymbol{\alpha}^T\underline{S}\boldsymbol{\beta}_0$ for any given $\boldsymbol{\alpha}$. $\boldsymbol{\alpha}_0$ and $\boldsymbol{\beta}_0$ are the actual distributions. If $p_i$ and $p_f$ are given as in the case of fermions, I can similarly define both $\alpha[p]$ and $\beta[q]$ and the two optimal distributions $\boldsymbol{\alpha}_0$ and $\boldsymbol{\beta}_0$ will extremize $\boldsymbol{\alpha}^T\underline{R}\boldsymbol{\beta}$.

As shown in [11] there is a stationary result for a system with $n$ possible paths. This extrema has $\boldsymbol{\alpha}_0$ parallel to $\underline{S}\boldsymbol{\beta}_0$, all components of $\boldsymbol{\alpha}_0$ are equal, and all components of $\boldsymbol{\beta}_0$ are equal. To get this solution, I assume that a given set of paths has the same probability to be observed independently of whether they are represented in terms of $p$ or $q$. This extrema has an analogy in the mini-max extrema of zero sum games [12]. Unlike a zero-sum game, however, the generalized form of Hamilton's principle may have solutions where individual elements of an optimal vector may be negative, positive, or anti-communitive. The element, e.g., $S_{12}$ may be large and $S_{12}\beta_2$ would seem to be a large potential positive payoff to $B$. $A$'s element $\alpha_1$, however, may be negative to create a potentially large loss to $B$ $\alpha_1 S_{12}\beta_2$. This concept of the mixed path allows $A$ or $B$ to select a negative value for an element of their mixed path in order to "negate" a large expected value of the action for the other experimenter (player). *A* or *B* may be attracted or repelled from a path because of the large expected value of the generalized action that the other "player" may get. The game analogy also suggests conceptual relations to competitive behavior in complex systems, biology, economics, etc. In fact, this view may provide a framework for a conceptual unification with these areas.

**The probability amplitude**

Even though $\underline{\alpha}_0$ is parallel to $\underline{S}\underline{\beta}_0$, $\underline{\alpha}_0$ is parallel to $\underline{\beta}_0$ only in the extraordinary case when $\underline{S}$ is diagonal. I may combine these two vectors using common parameters if they are expressed in a common basis. In the 2-D case, as shown in Figure 3, $\{\alpha_1, \beta_1\}$ is $\{|a|cos(\theta_1), |a|cos(\theta_1')\}$ and $\{\alpha_2, \beta_2\}$ is $\{|a|cos(\theta_2), |a|cos(\theta_2')\}$. The angles $\begin{pmatrix}\theta_1\\\theta_2\end{pmatrix}$ and $\begin{pmatrix}\theta_1'\\\theta_2'\end{pmatrix}$ specify the angle between $\begin{pmatrix}\alpha_o\\\beta_0\end{pmatrix}$ and the $\begin{pmatrix}x_1\\x_2\end{pmatrix}$ axes. If $x_1$ and $x_2$ are rotated so that $\theta_1 + \theta_1' = \pi/2$ and $\theta_2 + \theta_2' = \pi/2$ are satisfied, then I may use a complex representation: $\phi_1 = \alpha_1 + i\beta_1 = |a|cos\theta_1 + i|a|sin\theta_1 = |a|exp(i\theta_1)$ and $\phi_2 = |a|exp(i\theta_2)$. This same analysis is easily generalized to many paths, e.g., for three paths the two vectors with a given angle between them can be oriented with 3 angles in a 3 dimensional vector space that is fixed by three constraints between the corresponding angles (or direction cosines), $\theta_i + \theta_i' = \pi/2$, $i=1,2,3$. In *n* dimensions there are *n* paths and *n* constraints between corresponding angles. I can then find the total probability amplitude $K(q_i,t_i;q_f,t_f)$ as the sum over all $\phi_j$, i.e., $K(q_i,t_i;q_f,t_f) = \Sigma\phi_j = a\Sigma exp(i\theta_j)$ between the given endpoints. In relativistic quantum mechanics Lorentz invariance suggests that the probability amplitude for path *j* has $\theta_j=2\pi(S_j/h)$, where $S_j$ is a Lorentz invariant action for the system in question and *h* is Planck's constant [13]. In words, the phase angle of an ideal path is proportional to the action measured with respect to its limit of empirical existence (ontological uncertainty). The usual argument for the additivity of $S_j$ between consecutive endpoints implies the multiplying of *K*'s between consecutive endpoints implying the path integral formulation of quantum mechanics [13] and quantum field theory [9]. The normalization for bosons or fermions translating in space-time is written as $KK^*$. The probability amplitude for forward and backward paths naturally factorize for these mixed paths, suggesting the transactional interpretation of quantum mechanics [6]. *K* corresponds to the forward evolving offer wave and $K^*$ corresponds to the backward evolving confirmation wave and the paradoxes of quantum theory are resolved through the non-local and a-temporal transaction.

**Conclusion**

A new connection between classical mechanics and quantum mechanics is proposed. The quantum mechanical propagator is derived from a generalized form of Hamilton's principle. A system's non-local existence, a distribution functional over possible $p$ paths $\alpha[p(t)]$, a distribution functional over possible $q$ paths $\beta[q(t)]$, and a generalized action corresponding to a matrix of the action evaluated at all possible $p$ and $q$ are defined. The generalized Hamilton's principle is the extremization over all possible distributions of $\int\int_{p,q} Dq Dp\, \alpha[p]\, S[p,q]\, \beta[q]$ that is analogous to a zero sum game. The normalization of the distributions allows their values to be real numbers between *+1* and *−1* for particle translations and bosons; and Grassman numbers (between *+1* and *−1*) for fermions. The two optimal distributions are identified as the real and imaginary parts of the complex amplitude in a particular basis. In this theory I have used the effective limit $\delta q \to 0$, $\delta p \to \infty$ ; $\delta t \to 0$, $\delta E \to \infty$ (and vice versa). These limits, however, violate HUP in a finite universe and I have used them as a relative or effective limit. This approach to quantum theory that generalizes Hamilton's principle using non-local and a-temporal (empirical) existence, suggests many new possible approaches to formulating theory, e.g., one approach to quantum gravity is to use a general functional, $\alpha_0[p(t)]=a(S)$ and $\beta_0[q(t)]=b(S)$ with invariant $S$, and sum them between initial and final empirically non-existing space-time intervals suggested by a finite universe.

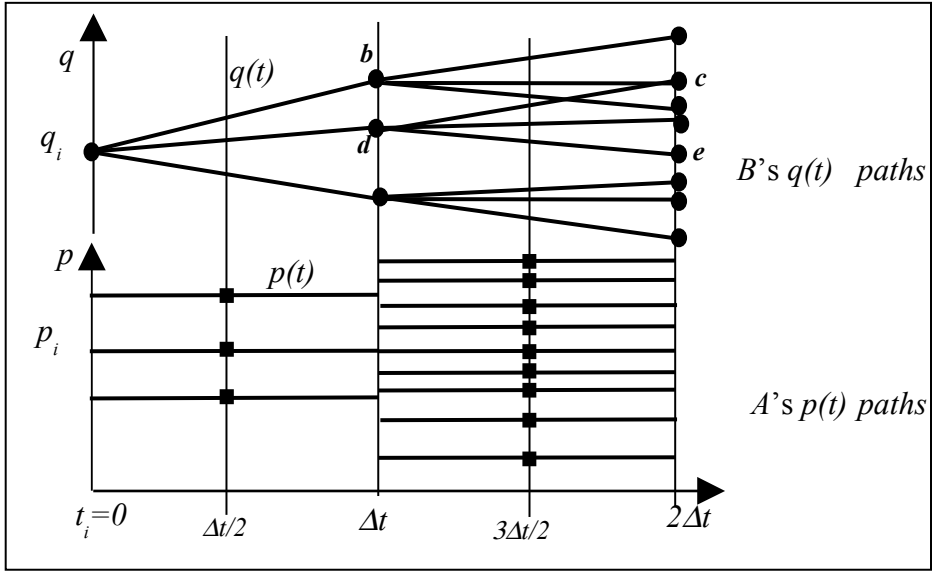

Figure 1. *q(t)* and *p(t)* evolution with a simplified uncertainty principle. *q* can be localized uniquely at any time but there are three possible *p*'s per definite *q* at each time. The number of possible paths increases geometrically with time (paths fan out). Each *q(t)* path that *B* could observe implies a corresponding *p(t)* path through Hamilton's equations (and vice versa for *A*). To avoid simultaneous knowledge of *p* and *q*, *A*'s and *B*'s sampling time are offset by $\Delta t/2$. The specific paths $\overline{q_i bc} \equiv \overline{q_i bcbc}$ or $\overline{q_i de} \equiv \overline{q_i bcde}$ are empirically equivalent, i.e., these are only counted as one path in the $\alpha$ and $\beta$ distributions. This shows that reversing the time in a forward moving path (between $t_i$ and $t_f$) is redundant. The possible paths divide into two classes of paths: 1) forward evolving and 2) backward evolving. The HUP may be inferred by generalizing to a continuous interval of points $\delta q$ and $\delta p$, such that $\delta q \delta p = h$. Note that HUP also implies that if the total interval of time (here $2\Delta t$) is less than the uncertainly $\delta t = h/E_{max}$, where $E_{max}$ is the total energy in the universe, then both forward and backward paths are possible because the time ordering of $t_i$ and $t_f$ is empirically non-existent.

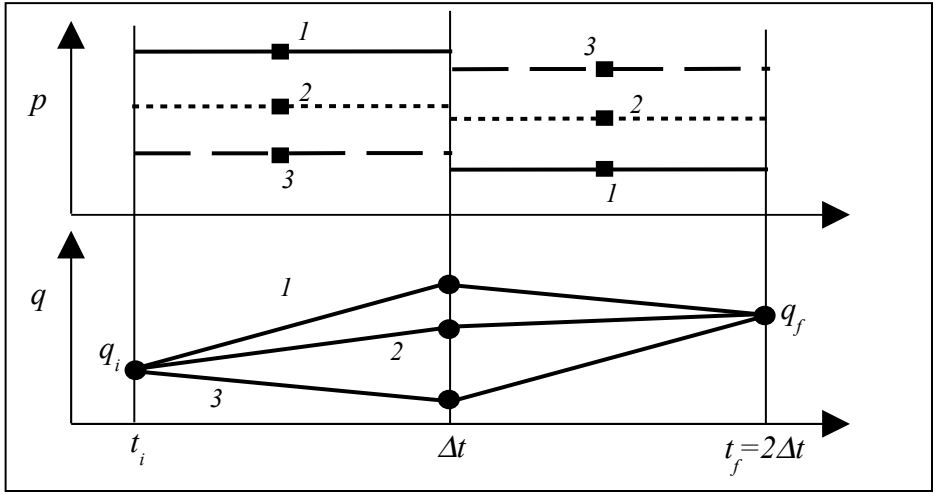

Figure 2. Five spatial-temporal points allow three possible paths for the simplified uncertainty principle used in Figure 1. The three $p_j(t)$ paths and the corresponding three $q_k(t)$ paths have mixed paths $\alpha=(\alpha_1, \alpha_2, \alpha_3)$, $\beta=(\beta_1, \beta_2, \beta_3)$, and an action matrix $S[p_j, q_k]=S_{jk}$. A and B attempt to extremize the generalized action $\alpha^T \underline{S} \beta$ by selecting optimal distributions $\alpha_0$ and $\beta_0$ such that the generalized action is extremized.

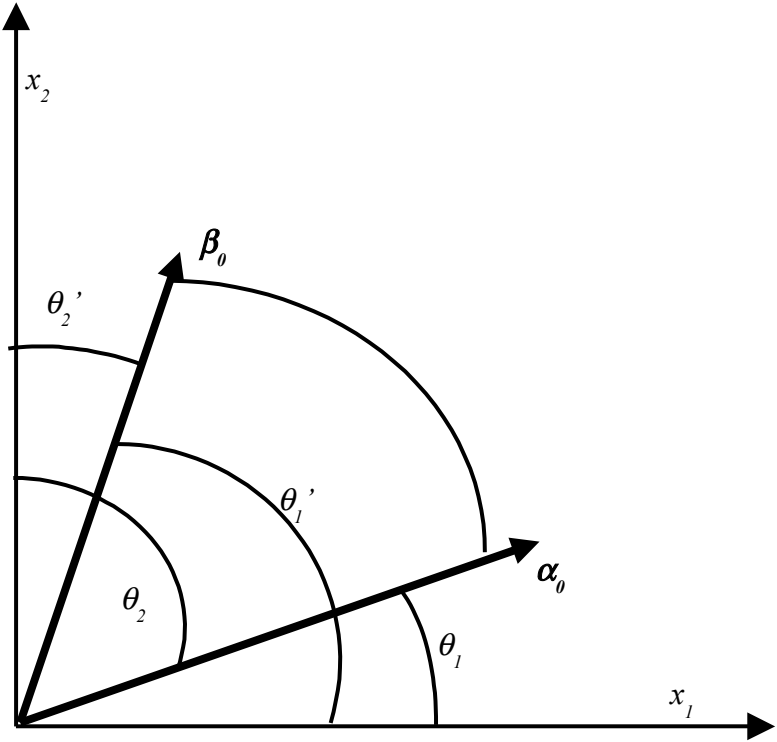

Figure 3. By combining the two optimal vectors $\alpha_0$ and $\beta_0$ in a common basis using the angles $\theta_1$, $\theta_1'$, $\theta_2$, and $\theta_2'$ where $\theta_1 + \theta_1' = \pi/2$ and $\theta_2 + \theta_2' = \pi/2$, I can then construct a complex quantity for path 1 and path 2, i.e., $\phi_1 = \alpha_1 + i\beta_1 = |a|cos(\theta_1) + i|a|sin(\theta_1) = |a|exp(i\theta_1)$ and $\phi_2 = \alpha_2 + i\beta_2 = |a|cos(\theta_2) + i|a|sin(\theta_2) = |a|exp(i\theta_2)$. $\phi_1$ and $\phi_2$ are the amplitudes for path *1* and *2* respectively.